\documentclass[12pt]{article}
\usepackage{fullpage}

\newcommand\pr{\prime}
\newcommand\be{\begin{equation}}
\newcommand\ee{\end{equation}}
\newcommand\bea{\begin{eqnarray}}
\newcommand\eea{\end{eqnarray}}
\newcommand\nn{\nonumber}

\newcommand\bdm{\begin{displaymath}}
\newcommand\edm{\end{displaymath}}

\def\pmb#1{\setbox0=\hbox{#1}%
  \kern-.025em\copy0\kern-\wd0
   \kern.05em\copy0\kern-\wd0
   \kern-.025em\raise.0433em\box0 }

\def\bk{\mathbf{k}}
\def\bx{\mathbf{x}}
\def\by{\mathbf{y}}

\newcommand\bvar{\mbox{\boldmath$\epsilon$}}
\def\mn{\mbox{\scriptsize $\mu\nu$}}

\begin{document}

\title{\large {\bf  Stochastic quantization of the linearized gravitational field\thanks{J.\ Math.\ Phys.\ {\bf 23}(1), January 1982, 132--137 \copyright  \,1982 American Institute of  Physics.}}}

\author{Mark Davidson\thanks{
Current Address: Spectel Research Corporation, 807 Rorke Way, Palo Alto, CA   94303
\newline  Email:  mdavid@spectelresearch.com, Web: www.spectelresearch.com}\\
\normalsize{\em Department of Physics, San Jose State University, San Jose, CA  95192}}

\date{ \normalsize (Received 25 August 1980; accepted for publication 31 October 1980)}

\maketitle

\noindent {\small Stochastic field equations for linearized gravity are presented.  The theory is compared with the usual quantum field theory and questions of Lorentz covariance are discussed.  The classical radiation approximation is also presented.
}


\section{Introduction}

Quantum systems which are described by a Schr\"odinger equation allow a stochastic interpretation.  The Fenyes-Nelson model \cite{fenyes1}--\cite{nelson2} and its generalization \cite{davidson1}--\cite{shucker1} provide such a stochastic description of quantum mechanics.  Several field theories have been considered in this context \cite{guerra1}--\cite{davidson4}, and the results have so far been of some interest.  In the present paper, the generalized Fenyes-Nelson model is applied to the weak field approximation of Einstein's general theory of relativity, the so-called linearized gravitational field \cite{einstein1}--\cite{weinberg1}.

It is worthwhile to attempt a stochastic model of quantum gravity for several reasons.  First, since the gravitational field in the classical theory has the interpretation of a metric tensor, it is disturbing that in the usual quantum formulation this geometric interpretation is completely obliterated because the field and its derivatives become abstract operators on a rigged Hilbert space.  Because of this difficulty it has become fashionable to play down the geometric role of gravity when dealing with quantization, as in the formulation of Weinberg \cite{weinberg1}.  A stochastic formulation  of the quantized gravitational field may revitalize the geometrical interpretation of quantum gravity.

Second, the stochastic formulation of quantum mechanics, at least in spirit, is a progeny of Einstein's profound discomfiture with the complementarity vision of Bohr.  Since Einstein's views were based at least in part on considerations of the general theory of relativity, it seems fitting to pursue a stochastic model of quantum gravity.

Third, the great successes which gauge theories have had in the theory of elementary particles suggest that fundamental efforts such as the stochastic reformulation of quantum mechanics be concentrated in this general area.

The linear theory of gravity has been selected for analysis because it avoids the extremely difficult problem of divergences in the full theory, and it is sufficiently simple to allow, perhaps, the beginning of a probabilistic geometric interpretation of quantum gravity.

In Sec.\ II the linearized Einstein field equations are briefly recounted.  In Sec.\ III the usual quantization procedure is outlined.  In IV the method of stochastic quantization is applied to the linear theory, and in V the curious random classical radiation approximation is presented.

\section{The linear field equations}

The Einstein field equations in the linear approximation may be written
\be
-\bar h_{\mu \nu,\alpha}^\alpha -n_{\mu \nu} \bar h_{\alpha \beta},^{\alpha \beta}  +\bar h_{\mu\alpha},^\alpha _\nu +\bar h_{\nu\alpha},^\alpha_\mu =16\pi GT_{\mu \nu},
\label{1}
\ee
where the notation of Ref.\ 13 is followed except that the gravitational constant $G$ is not set to unity.  Indices are raised and lowered via the Minkowski metric $n_{\mu \nu}$.  The full metric tensor is related to the above fields by
\bea
g_{\mu\nu}&=& n_{\mu \nu} +h_{\mu \nu},\label{2}\\
h_{\mn}&=& \bar h_{\mn} -\frac{1}{2} n_{\mn}\bar h^\alpha_\alpha,\label{3}\\
\bar h_{\mn} &=& h_{\mn} -\frac{1}{2} n_{\mn} h^\alpha_\alpha,
\label{4}
\eea
where $T_{\mn}$ is the energy-momentum tensor of all nongravitational sources.

Once the linear theory is solved, classically, the nonlinear corrections of the full theory may be included as a perturbation by introducing a suitable term in the energy-momentum tensor (Ref.\ 14, p.\ 165). This procedure has not worked so far in the quantum theory because of the celebrated divergences introduced by the nonlinear terms.

Weinberg has suggested that general relativity be treated as an ordinary field theory with the linear theory as the starting point.  This will be the approach taken here, although one of the primary motives for seeking a stochastic model of gravitation is to restore a geometric interpretation to the quantum theory.

The linear theory possesses a gauge invariance similar to electromagnetism (Refs.\ 13, p.\ 439; 14, p.\ 254).  The field equations are left invariant under the transformation
\be
h_{\mn}\to h_{\mn} -\partial {\cal C}_\mu /\partial x^\nu -\partial {\cal C}_\nu /\partial x^\mu,
\label{5}
\ee
where ${\cal C}^\mu$ is an arbitrary 4-vector field.  All observables are independent of the gauge, and so gauge conditions may be imposed by fiat in order to facilitate solution.  We shall work in the Lorentz gauge where one requires
\be
\bar h_{\mu ,\alpha}^\alpha =0.
\label{6}
\ee
The field equations become in this gauge
\be
-\partial_\alpha \partial^\alpha \bar h_{\mn} =16\pi GT_{\mn}.
\label{7}
\ee

The solutions to Eq.\ (\ref{7}) may be written as a sum of a retarded solution plus a free field solution,
\be
\bar h_{\mu\nu} =\bar h^R_{\mn} +\bar h^{\rm in}_{\mn},
\label{8}
\ee
where
\be
\bar h^R_{\mn} =\int \,4GT_{\mn} (\bx^\pr, t -|\bx -\bx^\pr|) /|\bx-x^\pr|d^3 x^\pr
\label{9}
\ee
and
\be
\partial_\alpha\partial^\alpha \bar h^{\rm in}_{\mn} =0.
\label{10}
\ee

The gauge condition [Eq.\ (\ref{6})] does not determine the gauge completely.   Any additional transformation such that
\be
\partial_\alpha \partial^\alpha {\cal C}_\mu =0,
\label{11}
\ee
will preserve the Lorentz condition.  This gauge transformation of the second kind can be used to make the free part of the field equation both transverse and traceless (Ref.\ 13, p.\ 946).  No generality is lost, therefore, by requiring
\bea
\bar h^{\rm in}_{o\nu}&=&0,\label{12}\\
\bar h^{\rm in}_{ii}&=& 0.
\label{13}
\eea

Periodic boundary conditions in the spatial coordinates will be imposed as an aid to quantization.  If the length associated with this periodicity is $L$, then wave numbers in a Fourier decomposition are restricted to have components which are integral multiples of $2\pi /L$.  Although the energy momentum tensor must also have this periodicity for consistency, no generality is really lost since $L$ will be taken to infinity eventually.  With these conditions, the Fourier decomposition of the free part of the gravitational field is
\be
h^{\rm in}_{ij} = \left(1/\sqrt 2 L^{3/2} \right) \sum\limits_{\lambda, \bk} \bvar_{ij} (\lambda, \bk) e^{i\bk\cdot \bx} {\cal Q} (\lambda, \bk, t),
\label{14}
\ee
where we have dropped the distinction between $\bar h^{\rm in}$ and $h^{\rm in}$, since in the transverse traceless gauge they are the same.  $\lambda$ denotes the polarization state and it can take on one of two values.  The $\bvar$'s must satisfy the conditions
\bea
&& \bvar_{ij} =\bvar_{ji},\quad k_i \bvar_{ij} =0, \quad \bvar_{ij} \bvar_{ij}=1, \quad \bvar_{ii}=0,\nn\\
&& \bvar_{ij} (\lambda)\,\bvar_{ij} (\lambda^\pr)= \delta_{\lambda, \lambda^\pr}, \quad \bvar_{ij} (\lambda , -\bk) = \bvar^{\bf *}_{ij} \, (\lambda, \bk).
\label{15}
\eea
The polarization tensors can be chosen real, and, with $\bk$ in the $z$ direction, they can be taken as
\bea
\bvar_{ij} (+) &=&\frac{1}{\sqrt 2} \,\left(\begin{array}{rrr}
1&0&0\\
0&-1& 0\\
0&0&0\end{array}
\right)\, ,\label{16}\\
\bvar_{ij}(\times)&=& \frac{1}{\sqrt 2} \, \left(\begin{array}{rrr}
0&1&0\\
1&0&0\\
0&0&0\end{array}
\right) \, ,
\label{17}
\eea
for the two independent states of polarization denoted by $+$ and $\times$.  The following expression is often useful:
\be
\sum\limits_{\lambda} \bvar_{ij} (\lambda) \bvar_{kl} (\lambda) =\frac{1}{2} \left( -\delta^{\rm tr}_{ij} \delta^{\rm tr}_{kl} +\delta^{\rm tr}_{ik} \delta^{\rm tr}_{jl} +\delta^{\rm tr}_{il} \delta^{\rm tr}_{jk}\right),
\label{18}
\ee
where
\be
\delta^{\rm tr}_{ij} =\delta_{ij} -k_i k_j/\bk^2.
\label{19}
\ee

The equations of motion for the field require that the ${\cal Q}$'s satisfy
\be
\ddot{\cal Q} +\bk^2 {\cal Q} =0,
\label{20}
\ee
so that a Lagrangian may be chosen of the form
\be
L=f \sum\limits_{\lambda,\bk} \left(\frac{1}{4} |\dot{\cal Q} |^2 -\frac{1}{4} \bk^2 |{\cal Q}|^2\right),
\label{21}
\ee
where $f$ is an as yet undetermined parameter.  In the sum over $k$ each ${\cal Q}$ appears twice because reality requires:
\be
{\cal Q}(\lambda, -\bk)= {\cal Q}^{\bf *}(\lambda, \bk).
\label{22}
\ee
The conjugate momenta are
\be
\frac{\partial L}{\partial {\rm Re} \dot{\cal Q}} = f\, {\rm Re}\dot{\cal Q} {\rm Re}P, \quad \frac{\partial L}{\partial {\rm Im}\dot{\cal Q}}= f{\rm Im} \dot{\cal Q}= {\rm Im} P,
\label{23}
\ee
and the Hamiltonian may be written as
\bea
H&=& \sum\limits_{\lambda, \bk} \left(\left| P\right|^2 /4f +\frac{1}{4} \bk^2 f\left|{\cal Q}\right|^2\right)\nn\\
&=&f\sum\limits_{\lambda, \bk} \left(\frac{1}{4} \left|\dot{\cal Q}\right|^2 +\frac{1}{4} \bk^2\left|{\cal Q}\right|^2\right).
\label{24}
\eea
This must now be compared to the actual energy of the gravitational field to set the scale for quantization.

The physical energy to be associated with a free gravitational field has been calculated (Ref.\ 13, p.\ 955).  It is given by the 00 component of the stress energy tensor:
\be
T_{\mn} =\frac{1}{32\pi G} \int_\Omega \left(\partial_\mu \bar h^{\rm in}_{ij}\right) \left(\partial_\nu \bar h^{\rm in}_{ij} \right) \,d^3 x,
\label{25}
\ee
where the integration is taken over a three-dimensional cube $(\Omega)$ of edge length $L$.  One finds
\bea
T_{00}&=& \left( 1/32\pi G\right) \int_\Omega \left(\bar h^{\rm in}_{ij} \right)^2 d^3 x\nn\\
&=& \left( 1/32 \pi G\right) \sum\limits_{\lambda,  \bk} \left(\frac{1}{4} \left|\dot{\cal Q}\right|^2 +\frac{1}{4} \bk^2\left|{\cal Q}\right|^2 \right),
\label{26}
\eea
so that we may make the identification
\be
f=1/32 \pi G,
\label{27}
\ee
and therefore,
\be
P=\left( 1/32 \pi G \right) \dot{\cal Q}.
\label{28}
\ee

\section{Quantization}

Quantization is achieved by imposing commutation rules as follows at equal times:
\bea
&&(1/32 \pi G)[\mbox{\bf Re}\dot{\cal Q} (\lambda, \bk), \mbox{\bf Re}{\cal Q} (\lambda^\pr, \bk^\pr)]\nn\\
&&\quad =-i\hbar \delta_{\lambda\lambda^\pr} (\delta_{ \bk,  \bk^\pr} +\delta_{ \bk, -  \bk^\pr}),\label{29}\\
&&(1/32 \pi G) [\mbox{\bf Im}\dot{\cal Q} (\lambda, \bk), \mbox{\bf Im}{\cal Q} (\lambda^\pr, \bk )]\nn\\
&&\quad =-i\hbar \delta_{\lambda \lambda^\pr} (\delta_{  \bk,   \bk^\pr} -\delta_{  \bk,- \bk^\pr}).
\label{30}
\eea
In terms of the fields, the equal time commutation rules become
\be
\left[ \bar h_{ij} (\bx ), \bar h_{kl} (y)\right] =- i\hbar \delta_{ijkl} (\bx-\by),
\label{31}
\ee
where
\be
\delta_{ijkl} (\bx-\by) =\int\,\frac{d^3k}{(2\pi)^3} \, e^{i\bk\cdot (\bx-\by)} \delta_{ijkl} (\bk),
\label{32}
\ee
and where
\be
\delta_{ijkl} (\bk)= \sum\limits_{\lambda} \,\bvar_{ij} (\lambda, \bk) \bvar_{kl} (\lambda, \bk)
\label{33}
\ee
is given by Eq.\ (\ref{18}).  The dynamical relation which fixes the quantum theory is
\be
\ddot{\cal Q} = -\bk^2{\cal Q}.
\label{34}
\ee
The field theory is equivalent to uncoupled harmonic oscillators.  In the ground state one finds
\be
\langle 0\left|\bar h^{\rm in}_{ij} (x) \bar h^{\rm in}_{kl} (y)\right|0\rangle = (32\pi G) i\hbar \int\,\frac{d^4k}{(2\pi)^4} \, \frac{\delta_{ijkl} (\bk)\,e^{ik^\mu(x_\mu-y_\mu)}}{-k^\mu k_\mu+i\bvar},\, \quad t_x>t_y.
\label{35}
\ee
Using the rule of Gaussian combinatorics,
\be
\langle 0\left|T\phi_1 (x_1) \times \dots\times \phi_n (x_n) \right|0\rangle =\sum\limits_\pi \langle 0\left| T\phi_1 (x_1) \phi_2(x_2)\right| 0\rangle \times \dots \times \langle 0 \left| T\phi_{n-1} (x_{n-1})  \phi_n (x_n)\right|0\rangle,
\label{36}
\ee
where $n$ is even, $T$ denotes time ordering, $\pi$ denotes a sum over distinct permutations, and the higher oder moments for the ground state of the quantum field may be calculated.  The spectral energy density for the field is found to be the same as for electromagnetism:
\be
\rho(\omega) =(\hbar /2\pi^2)\omega^3,
\label{37}
\ee
corresponding to an energy of $\frac{1}{2}\hbar \omega$ for each degree of freedom of the filed in the ground state.

\section{Stochastic quantization}

The linearized gravitational field as a quantum system may be reduced to an infinite dimensional Schr\"odinger equation of the form
\be
\left[\sum\limits_i \left( -\frac{1}{2} \hbar^2 \,\frac{\partial^2}{\partial{\cal Q}^2_i}\right) +V ({\cal Q}) \right] \psi =i\hbar \, \frac{\partial \psi}{\partial t}.
\label{38}
\ee
The dimensionality of this equation can be made finite by imposing a cutoff in momentum space.  Since much of the theory upon which stochastic quantization relies has been done for only finite dimensional systems, it is convenient to impose a momentum cutoff which may be taken to infinity in most expressions of interest.  This is not an important limitation and it can be expected as the theory of Markov fields advances it will eventually be eliminated. For the present, we shall assume a momentum cutoff.  Let us chose the following notation
\be
\psi =e^{R+iS}, \quad \Delta_{\cal Q} =\sum\limits_i \,\frac{\partial^2}{\partial {\cal Q}_i^2},
\label{39}
\ee
where $R$ and  $S$ are real functions.

Direct computation shows that the following equation is equivalent to Eq.\ (\ref{38}) so long as $R$ and $S$ have a first time derivative and a second ${\cal Q}$ derivative, and so long as $z$ has a nonzero real part
\be
\left[ -\frac{(z\hbar)^2}{2} \Delta_{\cal Q} +\left( V({\cal Q}) -\frac{\hbar^2}{2} \left(z^2 -1\right) \frac{\Delta_{\cal Q} e^R}{e^R}\right)\right] \,e^{R+iS/z}=i(z\hbar) \, \frac{\partial}{\partial t} \,e^{R+iS/z},
\label{40}
\ee
where $z$ is a constant which may be complex.

Suppose that $z$ is is purely imaginary, so that
\be
z=i|z|.
\label{41}
\ee
Then Eq.\ (\ref{40}) is still true if (\ref{38}) is true, but (\ref{40}) is only one real equation
\be
\left\{ \frac{\left(\left|z\right|\hbar\right)^2}{2} \Delta_{\cal Q} +\left( V+\frac{\hbar^2}{2} \left( 1+ \left|z\right|^2\right) \, \frac{\Delta_{\cal Q} e^R}{e^R} \right)\right\} \,e^{R+S/|z|} =-|z|\hbar \,\frac{\partial}{\partial t} \, e^{R+S/|z|}.
\label{42}
\ee
This equation must be supplemented by another real equation since the original Schr\"odinger equation contains two real equations.  Another equation may be generated by choosing
\be
z=-i|z|,
\label{43}
\ee
which leads to
\be
\left[ \frac{(|z|\hbar)^2}{2} \Delta_{\cal Q} +\left( V+\frac{\hbar^2}{2} \left(1+|z|^2\right) \,\frac{\Delta_{\cal Q} e^R}{e^R}\right) \right] e^{R-S/|z|}=|z|\hbar \frac{\partial}{\partial t} e^{R-S/|z|}.
\label{44}
\ee
Equations (\ref{42}) and (\ref{44}) taken together are equivalent to Schr\"odinger's equation by direct computation.

Consider the two real Eqs.\ (\ref{42}) and (\ref{44}).  They have the mathematical form of the heat equation.  Equations of this generic form are characteristic of a  certain type of diffusion problem called generalized Brownian motion or Ito processes.  In Ref.\ 4 equations of this form were derived within the context of diffusion theory.  The fact that Schr\"odinger's equation can be rewritten in the form (\ref{42}) and (\ref{44}) is the basis for the stochastic interpretation of quantum mechanics.

The stochastic interpretation rests on the hypothesis that all of the experimentally verifiable predictions of quantum mechanics may be deduced from Schr\"odinger's equation. This hypothesis shall be assumed true in this paper.

Stochastic quantization is achieved, following Nelson \cite{nelson1,nelson2} by associating with each coordinate ${\cal Q}$ a stochastic process which is defined by the stochastic differential equation:
\be
d{\cal Q}_i =b_i ({\cal Q}, t) +dW_i(t),
\label{45}
\ee
where the $W$'s are Wiener processes which satisfy
\be
E(dW_i dW_j) =2\nu \delta_{ij} dt.
\label{46}
\ee
Processes defined by Eq.\ (\ref{45}) go by the name: generalized Brownian motion, Ito processes, or multidimensional diffusion processes.  The mathematical background contained in Ref.~2 is sufficient for an understanding of stochastic quantization.  Other good sources for this subject are Refs.\ 15--22.  Since the formalism of forward and backward derivatives has been developed only by Nelson \cite{nelson1}, a careful reading of his book is essential to an understanding of stochastic quantization.  A method relying on an operator formalism, rather than the forward and backward time derivatives was presented in \cite{davidson1}, but it is completely equivalent to the Nelson procedure.

Many existence and uniqueness theorems for the processes defined by Eq.\ (\ref{45}) are presented in Refs.\ 2,  15--22.  The most important theorems show that if $b$ satisfies a global Lipschitz condition then Eq.\ (\ref{45}) has a solution which is a continuous Markov process and which is unique (for example, Ref.\ 2, page 43).

In Ref.\ 2 forward and backward time derivatives $D$ and $D_*$ are defined by
\bea
Df({\cal Q}, t) &=& \lim\limits_{h\to 0_+} \frac{1}{h} E(f({\cal Q} (t+h) , t+h)\nn\\
&&\quad -f({\cal Q} (t), t) |{\cal Q} (t) ={\cal Q} ),\label{47}\\
D_* f({\cal Q} , t) &=& \lim\limits_{h\to 0_+} \frac{1}{h} E(f({\cal Q}(t),t)\nn\\
&&\quad -f({\cal Q} (t-h) , t-h)|{\cal Q} (t) ={\cal Q}),
\label{48}
\eea
and independent of any dynamical assumption, these satisfy
\bea
D{\cal Q}_i &=& b_i,\label{49}\\
D_* {\cal Q}_i &=& b_{i_*},\label{50}\\
b_i -b_{i_*}&=& 2\nu \frac{\partial}{\partial {\cal Q}_i} \ln [\rho({\cal Q}, t)],\label{51}\\
D&=& \frac{\partial}{\partial t} +\sum\limits_i b_i \frac{\partial}{\partial {\cal Q}_i} +\nu \Delta_{\cal Q},\label{52}\\
D_*&=& \frac{\partial}{\partial t} +\sum\limits_i  b_{i_*} \frac{\partial}{\partial {\cal Q}_i} -\nu \Delta_{\cal Q}.
\label{53}
\eea

The dynamical assumption which leads to Schr\"odinger's equation is
\be
\frac{1}{2} (DD_*+ D_* D){\cal Q}_i +\frac{\beta}{8} (D-D_*)^2 {\cal Q}_i =-\frac{\partial}{\partial {\cal Q}_i} V,
\label{54}
\ee
where $\beta$ is a constant and where
\be
\nu =\hbar /2 \sqrt{(1-\beta/2)}.
\label{55}
\ee
The Schr\"odinger wave function is written in the form
\be
\psi=e^{R+izS_N},
\label{56}
\ee
where
\be
z=1/(1-\beta/2)^{1/2}
\label{57}
\ee
and
\be
b_i =2\nu \frac{\partial}{\partial {\cal Q}_i} (R+S_N).
\label{58}
\ee
It is straightforward to show that Eq.\ (\ref{54}) implies that (\ref{56}) satisfies (\ref{38}).

If $\nu$ is to be real, then we must demand the condition
\be
\beta<2.
\label{59}
\ee
It is possible, because of Eq.\ (\ref{55}), to choose any value of the diffusion parameter $\nu$ for a stochastic model of quantum mechanics.  In the limit when $\nu\to 0$, the model becomes deterministic and equivalent to Bohm's hidden variable thoery \cite{bohm1}, as was first pointed out in \cite{shucker1}.

In applying this formalism to linear gravity, we use the ${\cal Q}$'s in Eq.\ (\ref{14}), but mindful of the condition (\ref{22}) which relates ${\cal Q}$'s for antiparallel wave vectors.   If we define
\bea
b_{ij} &=& D\bar h^{\rm in}_{ij},\label{60}\\
b_{ij^*}&=& D_* \bar h^{\rm in}_{ij},
\label{61}
\eea
then with the aid of the operator
\be
\Delta^{ij} (\bx, t) = \frac{1}{(\sqrt 2) L^{3/2}}\,\sum\limits^{\lambda, \bk} \bvar^{ij} (\lambda, \bk) e^{i\bk\cdot \bx}\times \left( \frac{\partial}{\partial {\rm Re}{\cal Q}} \,+ i \, \frac{\partial}{\partial {\rm Im}{\cal Q}}\right),
\label{62}
\ee
the forward and backward time derivatives can be expressed as
\bea
D&=& \frac{\partial}{\partial t} +\int_\Omega b_{ij} \Delta^{ij} d^3 x +\nu \int_\Omega \Delta^{ij}\Delta_{ij} d^3 x,\label{63}\\
D_*&=& \frac{\partial}{\partial t} + \int_\Omega b_{ij^*} \Delta^{ij} d^3 x-\nu \int_\Omega \Delta^{ij}\Delta_{ij} d^3 x.
\label{64}
\eea
If we define a random field by
\be
W^{ij} (\bx, t) =\frac{1}{(\sqrt 2) L^{3/2}} \sum\limits_{\lambda, \bk} \bvar^{ij} (\lambda, \bk) e^{i\bk\cdot \bx} W_{\lambda, \bk} (t),
\label{65}
\ee
where the real and imaginary parts of $W_{\lambda, \bk}$ are Wiener processes which are independent of one another and which satisfy
\bea
&& E[d({\rm Re}W_{\lambda, \bk} )d({\rm Re}W_{\lambda^\pr, \bk^\pr})] =2\nu \delta_{\lambda \lambda^\pr} (\delta_{\bk,\bk^\pr} +\delta_{\bk, -\bk^\pr}) dt, \label{66}\\
&&E [d({\rm Im}W_{\lambda, \bk}) d ({\rm Im}W_{\lambda^\pr, \bk^\pr} )] = 2\nu \delta_{\lambda \lambda^\pr} (\delta_{\bk, \bk^\pr} -\delta_{\bk, -\bk^\pr}) dt,
\label{67}
\eea
then the stochastic differential equation may be written as
\be
d\bar h^{\rm in}_{ij} =b_{ij} dt +dW_{ij}.
\label{68}
\ee
With the gravitational field expressed as in Eq.\ (\ref{8}), and with the retarded solution still given by Eq.\ (\ref{9}), the field equations for the free part of the field become
\be
\left[\frac{1}{2} (DD_* +D_* D) +\frac{1}{8} \beta (D-D_*)^2 -\partial_l \partial^l\right] \bar h^{\rm in}_{ij}=0,
\label{69}
\ee
which is the stochastic version of Eq.\ (\ref{10}) in the transverse traceless gauge.

In order to solve these equations, Schr\"odinger's equation must first be solved in ${\cal Q}$ space.  Then the $b$'s are calculated using Eqs.\ (\ref{56}) and (\ref{58}).  Once the $b$'s are known, Eq.\ (\ref{68}) can in principle be solved.

In general, it is more difficult to solve the stochastic equations than to solve the usual quantum mechanical equations.  Even for the free field the stochastic processes for excited states are difficult to calculate.  So far, the stochastic method has proven useful in practical problems only for stationary state problems where considerable simplifications occur.  See, for example, Simon \cite{simon1} for a review of results in this area.  Although Simon does not explicitly make the connection with Nelson's theory, many of the methods he discusses may be considered as applications of the Fenyes-Nelson model to stationary state quantum systems.

We now illustrate the theory for the ground state field.  The solution to Schr\"odinger's equation in ${\cal Q}$  space for the ground state is
\be
\psi ({\cal Q}) =\prod\limits_{\lambda, \bk} \exp \left[ -|{\cal Q}(\lambda, \bk)|^2 \,\frac{\omega}{\hbar 128 \pi G}\right], \, \omega =|\bk|
\label{70}
\ee
up to a normalization constant.  The $b$'s are found from Eqs.\ (\ref{56}) and (\ref{58}) with $S_N =0$.  One finds for the $b$'s of Eq.\ (\ref{6})
\be
b^{ij} (x,t) =-2\nu /(\sqrt 2) L^{3/2} \times \sum\limits^{\lambda, \bk} \bvar^{ij} (\lambda, \bk) \,e^{i\bk\cdot\bx} \frac{\omega}{\hbar 32 \pi G} \, {\cal Q} (\lambda, \bk, t).
\label{71}
\ee
The stochastic differential equation becomes
\be
d\bar h^{ij} =b^{ij} dt +dW^{ij}.
\label{72}
\ee
Using the property of the $W^{ij}$ in Eq.\ (\ref{66}),
\be
E [dW^{ij} (x, t_x) dW^{kl} (y,t_y)] =2\nu dt\delta^{ijkl} (\bx-\by),
\label{73}
\ee
where the delta function is that of Eq.\ (\ref{32}), the stochastic equations may be integrated to yield
\be
E(\bar h^{ij} (x) \bar h^{kl} (y) ) =\hbar (16\pi G) \int\,\frac{d^3k}{(2\pi)^3} \, e^{i\bk\cdot (\bx -\by)}
\times e^{-(\nu /16 \pi G\hbar)|\bk||t_x -t_y|} \, \delta^{ijkl} (\bk)/|\bk|.
\label{74}
\ee
Since $b^{ij}$ of Eq.\ (\ref{71}) is linear in the ${\cal Q}$'s, the process turns out to be Gaussian with zero mean so the covariance (\ref{74}) determines all higher moments from the rule (\ref{36}).  The covariance may also be written as a four-dimensional integral
\be
E(\bar h^{ij} (x) \bar h^{kl} (y)) =(32 \pi G\hbar)\int \,\frac{d^4k}{(2\pi)^4} \, \frac{\delta^{ijkl}(\bk)}{k^2_0 +\bk^2} \, e^{i\bk\cdot (\bx-\by)}
\times e^{ik_0\nu /(\hbar 16 \pi G)|t_x-t_y|}.
\label{75}
\ee
In expressions (\ref{73})--(\ref{75}) the infinite volume limit has been taken.  Nelson's value for the diffusion parameter is
\be
\nu_N =16 \pi G\hbar.
\label{76}
\ee
Comparing Eq.\ (\ref{75}) with (\ref{35}), it may be shown that the stochastic covariance in (\ref{75}) may be obtained from the quantum covariance in (\ref{35}) by analytically continuing
\be
t_x\to -i [\nu /(\hbar 16 \pi G)]t_x, \quad t_y\to i[\nu /(\hbar 16 \pi G)]t_y.
\label{77}
\ee
This procedure of analytic continuation yields the Schwinger function.  Using the rule (\ref{36}), we obtain the general result:  The moments of the stochastic theory are equal to the Schwinger functions of the quantum theory with the times scaled by the factor $\nu(\hbar 16\pi G)$.  When Nelson's value for the diffusion parameter is chosen [Eq.\ (76)] the stochastic covariances become equal to the Schwinger functions.   This result is similar to the results obtained in scalar field theory [7, 9, 10] and in electromagnetism [8, 9, 11].

Examining Eq.\ (\ref{75}), we see that the covariance is not manifestly Lorentz covariant, even discounting for the fact that we have chosen a noncovariant gauge.  Lorentz covariance is violated by more than just a harmless gauge transformation in Eq.\ (\ref{75}).  This is a surprising and perhaps paradoxical result which has been known for some time \cite{guerra1}--\cite{davidson4}.  Despite this lack of manifest Lorentz covariance in the ground state, there is good reason to believe that the experimental predictions of the stochastic theory are consistent with special relativity and are in fact the same  as ordinary  quantum theory.  The argument is as follows.   Since for any real value of $\nu$ we have a stochastic model for a given solution to Schr\"odinger's equation, and since we believe that this equation contains all of the experimentally verifiable predictions of quantum mechanics, then there is reason to think that it is impossible to measure the diffusion parameter.  If this is true, then all of the experimentally measurable predictions of the theory (scattering cross sections, line spectra, etc.) must be constant when considered as analytic functions of $\nu$.  Because of this, we may consider analytic continuations to complex $\nu$ without affecting measurable predictions of the theory.  When the stochastic theory is continued to
\be
\nu=i\hbar (16 \pi G),
\label{78}
\ee
then one finds that the moments of the stochastic theory become the Green's functions of quantum field theory.  This result was also found for the scalar field \cite{davidson3} and the electromagnetic field \cite{davidson4}.  The stochastic theory can be expressed in a mathematical form which is identical to ordinary quantum theory when $\nu$ is continued to the value in Eq.\ (\ref{78}) (or its complex conjugate).  See, for example, Ref.\ 25 for a derivation of the operator formalism of quantum mechanics within this framework.  The methods of \cite{davidson5} generalized easily to the present theory in ${\cal Q}$ space.

Stochastic quantization in the ground state for a field theory leads to moments which are analytical continuations to imaginary times of quantum moments.  The stochastic interpretation suggests that a certain reality be attributed to the theory so continued.  It is interesting that imaginary time continuations have played a surprisingly important role in modern physics.  Euclidean field theory \cite{simon2} has led to advances in constructive field theory.  Complex manifold techniques \cite{lerner/sommers} have led to a deeper understanding of gauge theories. Bound state problems like the Bethe-Salpeter equations are often best solved by making a ``Wick rotation'' to imaginary times.  Perhaps the results of stochastic quantization provide an explanation for the usefulness of imaginary time continuations.

\section{The approximation of random classical radiation}

In electromagnetism it has proven interesting in several applications to approximate the ground state of the quantum field by a superposition of classical plane waves with radom phases.  This approximation has become known as random electrodynamics \cite{boyer1} and it has been found that the diffusion of charged particles in harmonic oscillator potentials and exposed to such radiation leads convincingly to Schr\"odinger's equation with the correct nonrelativistic  Lamb shift \cite{delapena}.  We present this random phase approximation for the linear gravity theory in the hope that it may find similar uses and also for comparison with the stochastic theory.

We first write the metric perturbations as general solutions to the free field equations.  We consider only the free field in the transverse traceless gauge
\bea
h^{ij} (\bx, t) &=& \frac{1}{(\sqrt 2) L^{3/2}} \,\sum\limits_{\lambda, \bk} \bvar^{ij} (\lambda, \bk)\,e^{i\bk\cdot \bx} {\cal Q} (\lambda, \bk, t),\label{79}\\
{\cal Q}(\lambda, \bk, t) &=& 32\pi G \hbar/ |\bk|\times \left\{ \cos \left[ \omega t +\theta_1 (\lambda, \bk)\right] +i\cos \left[\omega t +\theta_2(\lambda, \bk)\right]\right\}.
\label{80}
\eea
Reality demands
\be
\theta_1 (\lambda , -\bk) =\theta_1 (\lambda, \bk), \quad \theta_2 (\lambda , -\bk) =\theta_2 (\lambda , \bk) + \pi.
\label{81}
\ee
The $\theta$'s are all independent of one another except  for the conditions in Eq.\ (\ref{81}).  They are random phases which take on values fro 0 to $2\pi$. The averaging process is carried out by integrating over these phases.  One finds for the covariance
\be
E_\theta (h^{ij} (x) h^{kl}(y)) =16 \pi G\hbar \int \,\frac{d^3k}{(2\pi)^3} \, e^{i\bk\cdot(\bx-\by)} \cos \left(\omega\left(t_x -t_y\right)\right) \delta^{ijkl} (\bk)/|\bk|.
\label{82}
\ee
It is easy to show that this is equal to the symmetrized quantum expectation
\be
E_\theta (h^{ij} (x) h^{kl}(y)) =\langle 0 | \mbox{ Sym}(h^{ij} (x) h^{kl}(y)) |0\rangle,
\label{83}
\ee
where $\mbox{Sym}$ denotes the symmetrization operation
\be
\mbox{Sym}(\phi_1 (x_1) \times \dots \times \phi_n (x_n)) =\sum\limits_P \,\frac{1}{n!}\, \phi_1 (x_1) \times \dots \times \phi_n (x_n),
\label{84}
\ee
and where $P$ denotes a sum over all permutations of the arguments of the fields.

It can be shown that in the infinite volume limit the higher the moments of the random phase average satisfy the Gaussian combinatoric rule [Eq.\ (\ref{36})].  This is true despite the fact that the individual Fourier coefficients in Eq.\ (\ref{79}) are not distributed normally.  The reason is the central limit theorem, as the field is a sum of an infinite number of independent random variables.  Since the symmetrized quantum expectations also satisfy the Gaussian combinatoric rule, it follows that
\be
E_\theta (h^{i_1 j_1} (x_1) \times \dots \times h^{i_n j_n} (x_n)) =\langle 0|\mbox{Sym}(h^{i_1 j_1} (x_1) \times \dots \times h^{i_n j_n} (x_n))|0\rangle.
\label{85}
\ee
A more detailed derivation of this result has been given for electromagnetism by Boyer \cite{boyer1} whose analysis can be applied with little modification to the linear gravity theory to derive Eq.\ (\ref{85}).  It is important to realize that (\ref{85}) is not true for a one-dimensional oscillator or even for a finite dimensional oscillator.  It can only be derived in the present case in the infinite volume limit.  Thus the random phase approximation does not give a very detailed model of the quantum field accurate down to the level of a few normal modes.  It is unlikely, in the author's opinion, that a consistent interpretation of the full quantum field theory could be obtained from a classical random phase approximation for field theory.  This is in sharp distinction to the stochastic quantization theory of Sec.\ IV which allows a consistent reinterpretation of all quantum phenomena.  Still, the random phase model can be useful when considering the effects of vacuum fluctuations on matter.


\begin{thebibliography}{99}
\bibitem{fenyes1} I.\ Fenyes, Z. Phys. {\bf132}, 81--106 (1952).

\bibitem{nelson1} E.\ Nelson, {\em Dynamical Theories of Brownian Motion\/} (Princeton U.\ P., Princeton, N.\ J., 1967).

\bibitem{nelson2}  E. Nelson, Phys.\ Rev.\ {\bf 150}, 1079 (1966).

\bibitem{davidson1}  M.\ Davidson, Physica A {\bf 96}, 465--486 (1979).

\bibitem{davidson2} M.\ Davidson, Lett.\ Math.\ Phys. {\bf 3}, 271--277 (1979).

\bibitem{shucker1} D.\ Shucker, Lett.\ Math.\ Phys. {\bf 4}, 61--65 (1980).

\bibitem{guerra1} F.\ Guerra and P.\ Ruggiero, Phys.\ Rev.\ Lett. {\bf 31}, 1022 (1973).

\bibitem{guerra2}  F.\ Guerra and M.\ I.\ Loffredo, Lett.\ Nuovo Cimento {\bf 27}, 41--45 (1980).

\bibitem{moore1} S.\ M.\ Moore,  Found Phys. {\bf 9}, 237--259 (1979).

\bibitem{davidson3} M.\ Davidson, Lett.\ Math.\ Phys. {\bf 4}, 101--106 (1980).

\bibitem{davidson4}M.\ Davidson, ``Stochastic Quantization of the Electromagnetic Field,'' J.\ Math.\ Phys. {\bf 22}, 2588 (1981).

\bibitem{einstein1}  A.\ Einstein, {\em The Meaning of Relativity\/} (Princeton U.\ P., Princeton, N.\ J., 1979).

\bibitem{misner} C.\ W.\ Misner, K.\ S.\ Thorne, and J.\ A.\ Wheeler, {\em Gravitation\/} [Freeman, San Francisco, 1973).

\bibitem{weinberg1}S.\ Weinberg, {\em Gravitation and Cosmology\/} (Wiley, New York, 1972).

\bibitem{levy1} P.\ Levy, {\em Processus Stochastique et Mouvement Brownien\/} (Gauthier-Villars, Paris, 1965).

\bibitem{doob1} J.\ L.\ Doob, {\em Stochastic Processes\/} (Wiley, New York, 1953).

\bibitem{ito1} K.\ Ito, {\em Lectures on Stochastic Processes\/} (Tata, Bombay, 1961).

\bibitem{ito2} K.\ Ito and H.\ P.\ Mckean, Jr., {\em Diffusion Processes and Their Sample Paths\/} (Springer-Verlag, Berlin 1965).

\bibitem{feller1} W.\ Feller, {\em An Introduction to Probability Theory and Its Applications\/} (Wiley, New York, 1971).

\bibitem{breiman1} L.\ Breiman, {\em Probability\/} (Addison-Wesley, Reading, Mass., 1968).

\bibitem{dynkin1}  E.\ B.\ Dynkin, {\em Markov Processes\/} (Springer-Verlag, Berlin, 1965).

\bibitem{strook1} D.\ W.\ Strook and S.\ S.\ Varadhan, {\em Multidimensional Diffusion Processes\/} (Springer-Verlag, Berlin, 1979).

\bibitem{bohm1} D.\ Bohm, Phys.\ Rev.\ {\bf 85}, 180--193 (1952).

\bibitem{simon1} B.\ Simon, {\em Functional Integration and Quantum Physics\/} (Academic, New York, 1979).

\bibitem{davidson5} M.\ Davidson, Lett.\ Math.\ Phys. {\bf 3}, 367--376 (1979).

\bibitem{simon2} B.\ Simon, {\em The $P(\phi)_2$ Euclidean (Quantum) Field Theory\/} (Princeton U.\  P., Princeton, N.\ J., 1974).

\bibitem{lerner/sommers} D.\ E.\ Lerner and P.\ D.\ Sommers, {\em Complex Manifold Techniques in Theoretical Physics\/} (Pitman, London, 1979).

\bibitem{boyer1} T.\ H.\ Boyer, Phys.\ Rev.\ D {\bf 11}, 790--808, 809--830 (1975).

\bibitem{delapena}  L.\ de la Pe\~{n}a and A.\ M.\ Cetto, J.\ Math.\ Phys. {\bf 20}, 469--483 (1979).

\end{thebibliography}
\end{document}